\documentstyle[aps,preprint,prc]{revtex}
\tighten

\begin{document}
\draft

\title{Photon-Photon luminosities in relativistic heavy ion
collisions at LHC energies}
\author{
Kai Hencken \and
Dirk Trautmann
}
\address{
Institut f\"ur theoretische Physik der Universit\"at Basel,
Klingelbergstrasse 82, 4056 Basel, Switzerland
}
\author{
Gerhard Baur
}
\address{
Institut f\"ur Kernphysik (Theorie), Forschungszentrum J\"ulich,
52425 J\"ulich, Germany
}
\date{March 16, 1995}

\maketitle

\begin{abstract}
Effective $\gamma$--$\gamma$ luminosities are calculated for various
realistic hadron collider scenarios.  The main characteristics of
photon-photon processes at relativistic heavy-ion colliders are
established and compared to the corresponding
$\gamma$-$\gamma$-luminosities at $e^+$-$e^-$- and future Photon
Linear Colliders (PLC).  Higher order corrections as well as
inelastic processes are discussed.  It is concluded that feasible
high luminosity Ca-Ca collisions at the Large Hadron Collider (LHC)
are an interesting option for $\gamma$-$\gamma$ physics up to about
100 GeV $\gamma$-$\gamma$ CM energy.
\end{abstract}


\narrowtext

\section{Introduction}
\label{Sec:Intro}

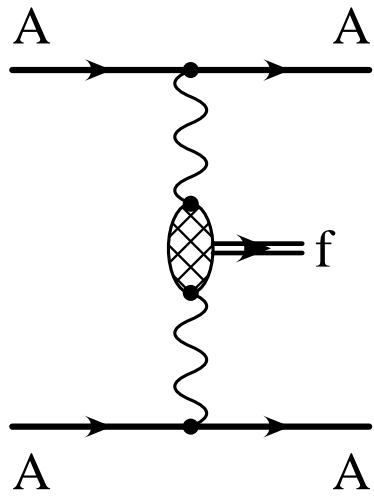
\begin{figure}
\caption{General form of an elastic photon-photon process.  f denotes
the system of the produced particles.}
\label{Fig:elastic}
\end{figure}

Due to the coherent action of $Z$ protons, strong electromagnetic
pulses of short duration are produced in relativistic heavy-ion
collisions.  Among other things, this can be useful for the study of
photon--photon collisions.  Up to now, photon--photon collisions have
been mainly studied at $e^+e^-$ colliders
\cite{gg92,ecfa86,budnev75}.  In contrast to the pointlike electrons
and positrons, nuclei are extended objects with an internal structure
and a size given by the radius $R \approx 1.2\ A^{1/3} $ fm, $A$ being the
mass number.  Also, the strong interaction between the nuclei in
central collisions leads to a lower cutoff of useful impact
parameters at $b \approx 2 R$.  In the last years the study of
$\gamma$-$\gamma$-interactions in heavy-ion collisions has been
pursued by many groups.  We refer to Refs.  \cite{baron94,vidovic93}
for a list of further references.  The spacelike virtual photon,
which is emitted by each nucleus, can be considered as quasireal.  An
exception is the production of $e^+e^-$ pairs \cite{hencken94}; in
this case, the mass of the electron $m_e$ is small compared to the
maximum value of $\sqrt{|q^2|}$, which is of the order of $1/R$.
Thus, generally, we can calculate $\gamma$-$\gamma$ cross sections in
RHI collisions in the factorized form (see also Fig.
\ref{Fig:elastic})
\begin{equation}
\frac{d\sigma}{d\Omega}\left( A A \rightarrow A A X \right) =
  \int dW
  \frac{dL_{\gamma\gamma}}{dW}
  \frac{d\sigma}{d\Omega} \left(\gamma\gamma \rightarrow X\right),
\label{Eq:Lum}
\end{equation}
where $W$ is the invariant mass of the photon-photon system.  The
effective $\gamma$-$\gamma$-luminosity $d\tilde L_{\gamma\gamma} /
dW$ is given by
\begin{equation}
\frac{d\tilde L_{\gamma\gamma}}{dW} =
\frac{dL_{\gamma\gamma}}{dW} L_{AA}
\end{equation}
where $L_{AA}$ is the $A$--$A$ luminosity.  Due to the coherence of
the photon emission, there is a $Z^4$ factor for the
$\gamma$-$\gamma$ cross section.  This factor is partly compensated
by the lower luminosity achieved for the heavy ions, as compared to
the $p$--$p$ case.

In Sec.  \ref{Sec:Luminosity} effective $\gamma$-$\gamma$
luminosities are compared with each other for various collider types.
We use the luminosities of the heavy-ion beams from a recent study by
Eggert and Morsch \cite{eggert94,eggertPC}.  Unrestricted, as well as
restricted (in the rapidity $Y$) $\gamma$-$\gamma$ luminosities are
discussed.  This extends the work of Ref.~\cite{baron94}.  In Sec.
\ref{Sec:Higher} we study the influence of higher order processes and
the effects of inelastic processes.  Inelastic processes are those,
where the emission of the equivalent photon leads to a nuclear
transition.  Of special interest is the case, where the nucleus makes
a transition to the giant dipole resonance (GDR), a very collective
nuclear mode.  The study of such kind of effects is important for a
detailed planning and optimization of the experiments, see e.  g.
\cite{sadovsky94,ALICE94}.  A more qualitative comparison to the case
of $\gamma$-$\gamma$--physics in $p$-$p$ collisions is also made.
Our conclusions are given in Sec.  \ref{Sec:Conclusions}.  In the
appendix we give a short derivation of the inelastic double
equivalent photon approximation (DEPA) and apply it to the nuclear
GDR excitation.

\section{Effective $\gamma$--$\gamma$-luminosities for hadron-
 and \lowercase{$e^+e^-$}-colliders}
\label{Sec:Luminosity}

In this section we are going to compare the effective
$\gamma$-$\gamma$ luminosities for Ca-Ca and Pb-Pb beams at the LHC
with those of other colliders.  We want to study whether the
heavy-ion beams can compete with other colliders.

We calculate the $\gamma$-$\gamma$ luminosities in the heavy-ion case
using the semiclassical impact-parameter-dependent form of the photon
density, as described in \cite{baur90,baur93}.  The $b$-dependent
equivalent photon number is given by
\begin{equation}
N(\omega,b) = \frac{Z^2 \alpha}{\pi^2}
\left(\frac{\omega^2}{\gamma^2}\right) K_1^2 \left(\frac{\omega
b}{\gamma}\right),
\end{equation}
where $K_1$ is the modified Bessel function of the second kind.  The
total unrestricted $\gamma$-$\gamma$ luminosity is then given by
\cite{baur90,cahn90}
\begin{eqnarray}
dL/dW &=& \frac{4 \pi}{W} \int dY \int_{R_1}^\infty b_1 db_1
\int_{R_2}^\infty b_2 db_2
\int_0^{2\pi} d\phi N( \frac{W}{2} e^{Y}, b_1)
N( \frac{W}{2} e^{-Y}, b_2) \nonumber\\
&&\Theta(b_1^2+b_2^2-b_1 b_2 \cos(\phi) - (R_1 + R_2)),
\end{eqnarray}
where we have introduced a cutoff in the impact parameter, in order
to discriminate for central collisions, where both ions interact
strongly with each other.  In order to have a clean signal, we drop
these events.  Due to this cutoff, we do not need a form factor
explicitly.  For $p$-$p$ and $e^+$-$e^-$ collisions we use the
formulae from the literature \cite{ohnemus93,ecfa86} (For the $p$-$p$
case only the elastic part of the luminosity is used.)
\begin{eqnarray}
dL/dW (pp) &=& \frac{2 W}{s} \frac{1}{\tau}
\left(\frac{\alpha}{\pi} \right)^2 \frac{2}{3} \log^3
\left(\frac{1}{\tau}\right)\\
dL/dW (e^+ e^-) &=& \left( \frac{\alpha}{2\pi} \log\left(\frac{s}
{4 m_e^2} \right)\right)^2 \frac{1}{\tau} \left[ \left(2 + \tau
\right)^2 \log \left(\frac{1}{\tau} \right) - 2 (1-\tau) (3+\tau)
\right],
\end{eqnarray}
with $\tau=W^2/s$ and $s=(P_1+P_2)^2$.  In the $e^+$-$e^-$ case the
formula also includes correction due to the collisions, where the
lepton can loose an appreciable part of its energy.  For the next
generation of photon-photon colliders, which are based on the Compton
backscattering of an electron beam in an intense laser beam
\cite{ginzburg94}, we use the results of Fig.~9 of
Ref.~\cite{barden94}.  Here we use the result for the ``conversion
distance'' $z=0$cm, where the photon luminosity is maximal.

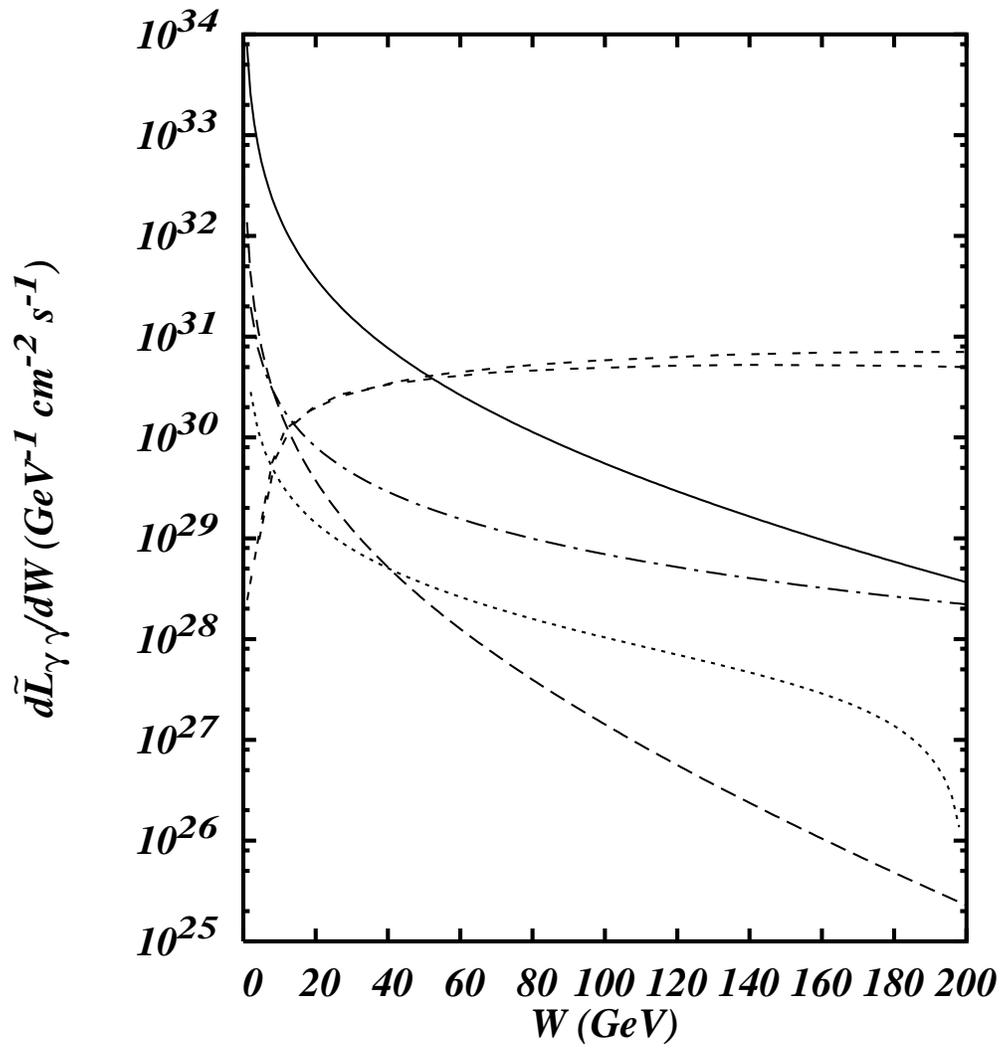
\begin{figure}
\caption{Comparison of the photon-photon effective luminosity
$d\tilde L_{\gamma\gamma}/dW$ as a function of the invariant mass $W$
of the produced system.  Shown are the results for hadron collider
using Ca-Ca (solid line), Pb-Pb (dashed line) and $p$-$p$ collisions
(dot-dashed line) together with those of $e^+$-$e^-$ (dotted line)
and $\gamma$-$\gamma$ colliders (two-dotted lines) for two different
polarizations.  See text and Table~\protect\ref{Tab:Collider} for the
parameter used for the different accelerators.}
\label{Fig:Lum_el}
\end{figure}

In Fig.~\ref{Fig:Lum_el} we compare the unrestricted luminosities of
these colliders.  The different beam luminosities and energies used
in the comparison are summarized in Table~\ref{Tab:Collider}.  We see
that the spectrum for $p$-$p$ collisions is much harder for higher
invariant masses than that for heavy ions.  This is due to the
smaller radius of the proton.  For small values of $W$ the different
luminosities are mainly proportional to the beam luminosity and
$Z^4$.  For small enough $W$, the $\gamma$ dependence is
given by \cite{baur88,bertulani88}
\begin{equation}
dL/dW = \left(\frac{Z^2 \alpha}{\pi}\right)^2 \frac{32}{3 W} \ln^3
\left(\frac{2 \gamma}{W R}\right).
\end{equation}
For the LHC, we see that the Ca-Ca collisions seem to be the desired
choice, whereas the Pb-Pb case is even lower than the $p$-$p$ case
for higher invariant masses.

The planned PLC would be superior, as a dedicated photon-photon
machine, for $W > 40 GeV$; its energy range exceeds what will be
possible with the LHC.  Please note that the photon-spectrum shown in
the figure is only the low energy part of it.  Most photons are
produced with energies in the TeV region.  But whereas the LHC is
accepted now and will be in operation in 2004 or 2008,
$\gamma$-$\gamma$ colliders are only in a planning stage.  Therefore,
we think that photon-photon physics at the LHC is an interesting
option, extending the range of the invariant mass to energies not
possible today.

As already discussed in \cite{baron94}, the unrestricted luminosity
is not the one relevant directly for experiments.  Depending on the
detectors used, only a part of all photon-photon processes may be
detected.  Which part of the processes are really observed, depends
on the detector system and on the produced particles.  The importance
of such a cutoff can be demonstrated, by using a simple formula for a
cutoff \cite{ecfa86}
\begin{equation}
 \left| p_{\gamma\gamma} \right| < f W.
\end{equation}
Setting $f=1$, we get a restriction on the $Y$ range with $-Y_0 < Y
< Y_0$ and
\begin{equation}
Y_0 =  \ln(1+\sqrt{2}).
\end{equation}

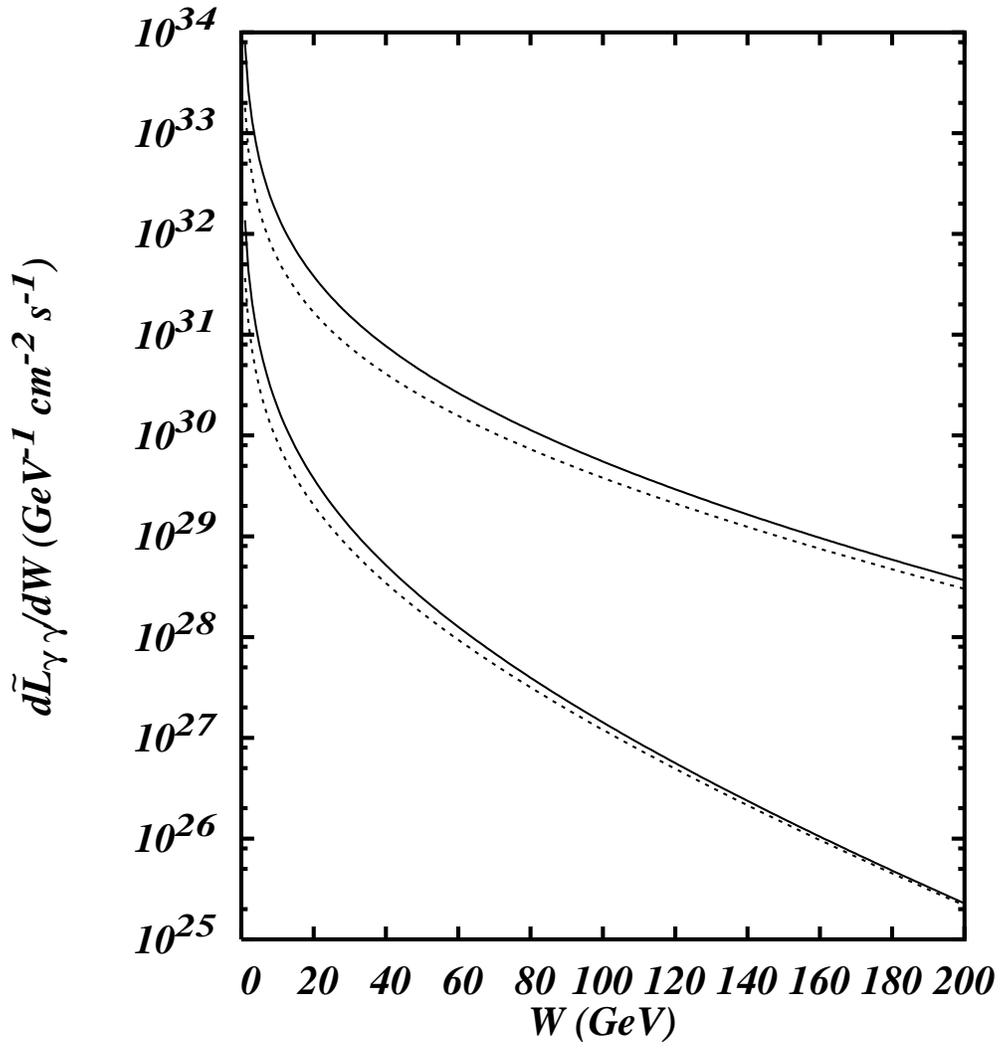
\begin{figure}
\caption{Comparison of the unrestricted (solid line) and restricted
effective luminosities (dotted line).  Shown are the results for
Ca-Ca collisions (upper curve) and Pb-Pb collisions (lower curve).
The Collider parameters used are shown in
Table~\protect\ref{Tab:Collider}.}
\label{Fig:Lum_restricted}
\end{figure}

In Fig.~\ref{Fig:Lum_restricted} we compare the restricted
luminosities with the unrestricted ones for Ca-Ca and Pb-Pb.  We see
that they are reduced for small invariant mass as expected
\cite{baron94}.

Due to the factorization property of Eq.~(\ref{Eq:Lum}), these
luminosities can now be used easily in order to calculate the
different $\gamma$-$\gamma$ processes, leading to the production of
resonances and continuum states.  For this we need the cross sections
for the photon-photon subprocess.  We refer here to the large
literature on this subject, especially the production of heavy quark
systems, the production of the Higgs boson and of supersymmetric
particles \cite{sadovsky94,ohnemus93,drees94}.

\section{Higher order and inelastic processes}
\label{Sec:Higher}

In contrast to the electron and positrons, nuclei are extended
objects with a rich excitation spectrum.  Now in connection with the
planning of the experiments, the question arises how this nuclear
structure influences the $\gamma$-$\gamma$ spectrum.  In Section
\ref{Sec:Luminosity} the effect of the finite size of the nucleus is
fully accounted for by the introduction of the cutoff $b_{min} = 2 R$
($b_1>R$, $b_2>R$) in impact parameter space, i.e., the elastic
charge form-factor of the nucleus is included.

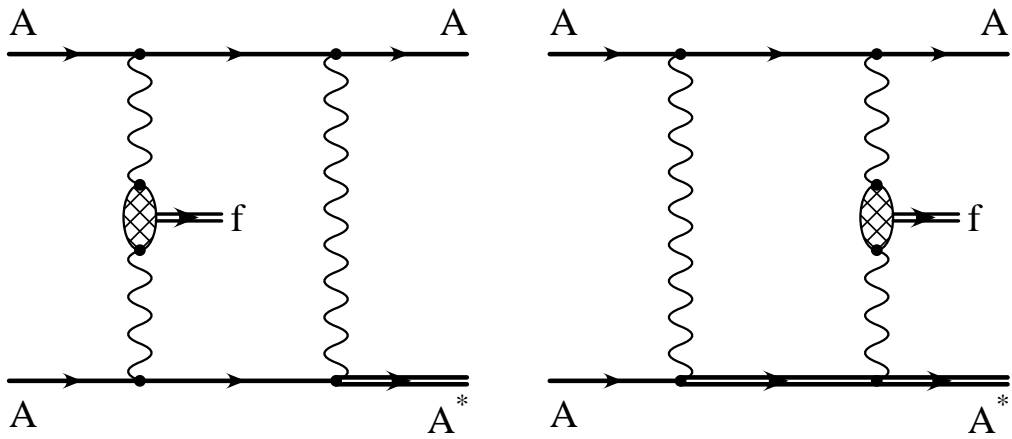
\begin{figure}
\caption{The two processes contributing to the higher order process,
where the $\gamma$-$\gamma$ process accompanied by a nuclear
excitation of one of the nuclei.}
\label{Fig:GDR_coh}
\end{figure}

In the following we consider two different effects, which lead to
excited nuclear states.  Firstly a higher order electromagnetic
process where there is, in addition to the $\gamma$-$\gamma$ process,
an inelastic $\gamma A$ interaction.  The process is shown in
Fig.~\ref{Fig:GDR_coh}.  The two different processes can be seen as
independent, if we assume, that the elastic form factor, which is
determined by the size of the system, remains nearly the same.  Then
it is simplest to adopt the semiclassical impact parameter approach.
The probability for a $\gamma\gamma \rightarrow f$ process and the
$\gamma A \rightarrow A^*$ process is given by the product
\begin{equation}
P_{fA^*}(b) = P_{\gamma\gamma \rightarrow f} (b) \ P_{\gamma A
\rightarrow A^*} (b).
\end{equation}
Integrating from $b=2R$ up to infinity, we obtain for the cross
section for $\gamma\gamma \rightarrow f$ fusion accompanied by
a specific $\gamma A \rightarrow A^*$ interaction
\begin{equation}
\sigma_{fA^*} = 2 \pi \int_{2R}^{\infty} b db \
  P_{\gamma\gamma \rightarrow f} (b)
  \ P_{\gamma A \rightarrow A^*} (b).
\label{Eq:sigmafA*}
\end{equation}

Since $\sum_{A^*} P_{\gamma A \rightarrow A^*} = 1$ (where the sum
over $A^*$ includes the ground state also), we have
\begin{equation}
\sum_{A^*} \sigma_{f A^*} = \sigma_{f} = 2 \pi
\int_{2R}^{\infty} b db P_{\gamma \gamma \rightarrow f} (b).
\label{Eq:sigmaf}
\end{equation}
$\sigma_f$ is the cross section of the $\gamma\gamma\rightarrow f$
process alone, without any higher order processes, which was
calculated up to now.  We define $\sigma_{f,A}$ as the cross section
for a $\gamma\gamma \rightarrow f$ process, not accompanied by a
$\gamma A$ interaction, and $\sigma_{f,A^*}$, where such an
interaction takes place.  An estimate can be given using the
following argument: Usually the integrand will be peaked at $b
\approx b_{min} = 2 R$, and we approximate therefore
\begin{equation}
\sigma_{f,A} \approx 2 \pi P_{A}(2R) \int_{2R}^{\infty} b db \
P_{\gamma\gamma\rightarrow f}(b) = 2 \pi P_{A}(2R)
\sigma_{\gamma\gamma\rightarrow f},
\end{equation}
and
\begin{equation}
\sigma_{f,A^*} \approx 2 \pi P_{A^*}(2R) \int_{2R}^{\infty} b db \
P_{\gamma\gamma\rightarrow f}(b) = 2 \pi P_{A^*}(2R)
\sigma_{\gamma\gamma\rightarrow f},
\label{Eq:GDR_el}
\end{equation}
where $P_{A} + P_{A^*} = 1$ and $P_{A^*} = \sum_{A^* \neq A}
P_{\gamma A \rightarrow A^*}$.

Impact parameter dependent probabilities were calculated in
\cite{baron94,bertulani88,vidovic94}.  They depend strongly on $A$.
Especially important is the excitation of the giant dipole resonance
(GDR), which is assumed to be located at
\begin{equation}
E_{GDR} = 80 MeV \frac{1}{A^{1/3}}.
\label{Eq:PGDR1}
\end{equation}
The excitation probability can be calculated, assuming that it has
zero decay width and that it exhausts the Thomas-Reiche-Kuhn (TRK)
sum rule.  As the probability for this excitation becomes large for
Pb-Pb-collisions, we include in our calculation the effects of
multiphonon excitation.  The sum over all these excitations is given
by (see Eq.~(3.2.4) of \cite{bertulani88})
\begin{eqnarray}
P_{GDR}(b) &=& 1 - \exp\left(-  \frac{Z^2 \alpha}{E_{GDR} \pi^2}
\frac{1}{b^2} \int_0^\infty \frac{d\omega}{\omega}
\sigma_{\gamma A}(\omega) \right)\nonumber\\
&\approx&1 - \exp\left(- \frac{Z^2 \alpha}{E_{GDR} \pi^2}
\frac{1}{b^2} 60 \frac{NZ}{A} \mbox{MeV}\  mb\right).
\label{Eq:PGDR2}
\end{eqnarray}

This gives an excitation probability of about 40\% for Pb and of
about 1\% for Ca at $b=2R$.  As each of the ions can be excited, we
get a total probability for the excitation of at least one of them of
about 2\% for Ca and of about 65\% for Pb.  With less probability the
quasideuteron region, the nucleon resonances and also the nuclear
continuum are excited.  While these effects are appreciable for the
Pb-Pb case, the effects are only of minor importance for lighter
systems like Ca-Ca, due to the strong $Z$ dependence.

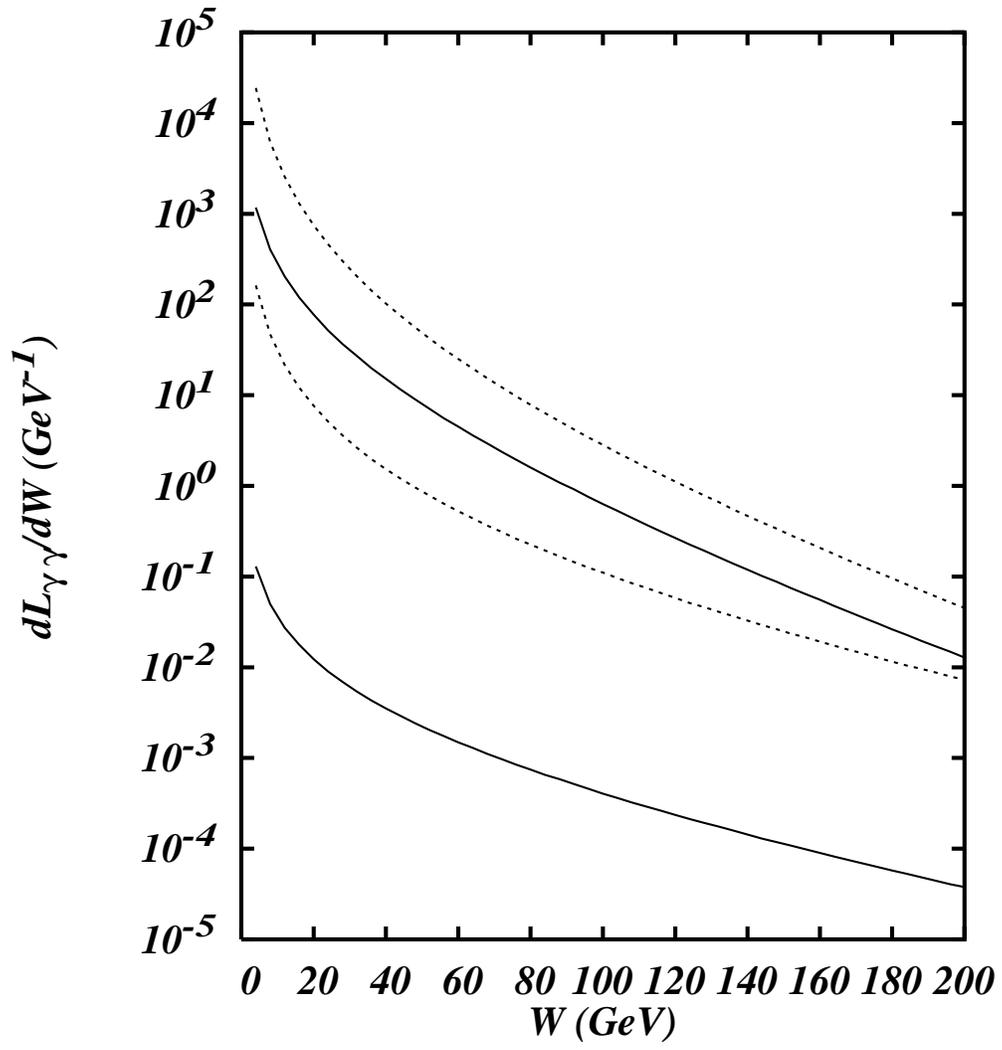
\begin{figure}
\caption{Comparison of the total $\gamma$-$\gamma$ luminosity (dotted
line, Eq.~(\protect\ref{Eq:sigmaf})), see Eq.~(\protect\ref{Eq:Lum}),
with the one, where it is accompanied by GDR excitation of one of the
ions (solid line, Eq.~(\protect\ref{Eq:sigmafA*}), where $A^*$ is the
GDR).  Shown are the results for Pb-Pb (upper curves) and Ca-Ca
(lower curves).}
\label{Fig:Lum_GDR_coh}
\end{figure}

A detailed calculation of the cross section using the $b$ dependent
probability as given by Eqs.~(\ref{Eq:PGDR1}) and~(\ref{Eq:PGDR2})
and integrating numerically over the impact parameter gives the
results of Fig.~\ref{Fig:Lum_GDR_coh}.  These are in qualitative
agreement with Eq.~(\ref{Eq:GDR_el}).  Whereas Eq.~(\ref{Eq:GDR_el})
is too high for small invariant masses, it gives the right order of
magnitude for large ones.  This is due to the fact, that processes
with small invariant mass are possible at larger impact parameter,
whereas high invariant masses are possible only for close collisions.

Most frequently the excitation of the GDR state is followed by
neutron evaporation.  This leads to a change of the mass to charge
ratio of the nucleus and the fragments will be lost from the
circulating beam and can possibly be detected in a zero degree
calorimeter (ZDC).

In addition to the elastic EPA of Sec.~\ref{Sec:Luminosity}, where
the nucleus remains in its ground state during the photon emission
process, there are also elastic-inelastic and even
inelastic-inelastic processes, where the emission of the photon leads
to the excitation of the nucleus.  The corresponding situation was
studied by Ohnemus et al.  \cite{ohnemus93} for $p$-$p$ collisions,
who found that inelastic processes are a non-negligible contribution in
this case.  Let us say a few words about the qualitative differences
between the $p$-$p$ and $A$-$A$ case.  A typical inelastic-elastic
process is shown in Fig.~\ref{Fig:IEPA_GDR}.  In the appendix, we
give a short derivation of the generalization of the elastic EPA to
inelastic processes.  On the one hand, we get a modification of the
formula due to the different vertex factor, on the other hand, the
minimal $q^2$ is modified as we have an inelastic process.  For the
most important case of the GDR excitation the spectrum of the
equivalent photons is also given in the appendix.  As demonstrated
there, the spectrum is much more sensitive to the detailed form of
the form factor in the inelastic case as in the elastic case.
Therefore we have used two different form factors, a dipole form
factor given by
\begin{equation}
	F(k^2) = \frac{\Lambda^4}{\left(\Lambda^2 + k^2\right)^2},
\end{equation}
and a Gaussian form factor
\begin{equation}
	F(k^2) = \exp\left(- \frac{k^2}{\lambda^2}\right).
\end{equation}
Integrating over $\omega$, we get the luminosities for the
elastic-inelastic processes as shown in Figs.~\ref{Fig:Lum_IEPA_Pb}
and~\ref{Fig:Lum_IEPA_Ca}.  The contribution of the
elastic-inelastic processes are about 1\%, so the detailed form of
the form factor is not important in this case.

Inelastic processes are only a small correction.  This can be
understood from the fact, that the electromagnetic transition current
remains finite for $q^2 \rightarrow 0$ only in the elastic case.  For
inelastic processes it vanishes at least with $q$, compensating
therefore the enhancement of the process due to the small $q^2$ in
the photon propagator.

Another process, which has to be considered, is the incoherent
scattering of an individual proton instead of the whole nucleus (see
Fig.~\ref{Fig:Dis}), where $q^2$ is larger than $1/R^2$.  It seems
interesting to compare the elastic and inelastic contributions to the
equivalent photon spectrum in the $p$-$p$ and $A$-$A$-cases (see also
\cite{ohnemus93}).  We consider a proton as being made up of $u$ and
$d$ quarks, with charges $2/3$ and $-1/3$, respectively.  A nucleus
is made up of $Z$ protons, which are then made up of quarks.  Whereas
the charge of the proton is comparable to the charge of its
individual constituents, the charge of the nucleus is much larger
than the charge of its constituents, if $Z \gg 1$.  Thus for nuclei,
the equivalent photon spectrum is dominated strongly by the coherent
component, where $q^2 \ll 1/R^2$, whereas for protons the
individual, incoherent quark contributions are comparable, or even
larger, than the elastic component.  This is in accordance with the
quantitative results of Ref.~\cite{ohnemus93}.

It has been suggested \cite{eggertPC} to use
$\mu^+\mu^-$-pair-production as a luminosity monitor for the $p$-$p$
and the $A$-$A$ collider.  The $\gamma$-$\gamma$ mechanism dominates
for collisions with no strong interactions between the protons.  This
condition can be assured experimentally.  In addition to the elastic
and incoherent contributions, also the transition to nucleon
resonances, notably the $\Delta$, have to be taken into account.
Since the relevant electromagnetic form factors are known, a reliable
calculation of the $\mu^+\mu^-$ production via the $\gamma$-$\gamma$
mechanism should be feasible.  The same formalism, which was used
here for the GDR excitation of a nucleus, can also be used for this
case.  This is the subject of future work, beyond the scope of the
present paper.

\section{Conclusions}
\label{Sec:Conclusions}

The possibility to do photon-photon physics with heavy ions at LHC
energies was scrutinized.  We calculate unrestricted and restricted
luminosities for Pb and Ca.  For large invariant masses heavy ions
notably Ca-Ca, with its relatively high luminosity, compare very
favorably with LEP200, e.g..  For very high invariant masses, $W
\gtrsim 200 $GeV, the $p$-$p$ option becomes better; this is
essentially due to the harder form factor of the proton as compared
to the softer nuclear form factor.  In the far future a dedicated
photon-photon collider (like PLC) will have a larger
$\gamma$-$\gamma$ luminosity than the $AA$ options over most of the
interesting invariant mass regions.

Also with the aim to help planning experiments, higher order and
other possibly important correction factors were investigated.  We
find that for very high $Z$, like Pb-Pb, most of the
$\gamma$-$\gamma$ collisions are accompanied by a nuclear excitation,
predominantly GDR excitation.  For the Ca-Ca case such effects are
almost negligible.  We show that in contrast to the $p$-$p$ case
$\gamma$-$\gamma$ events from inelastic photon emission are of minor
importance for $AA$ collisions.  We conclude that $\gamma$-$\gamma$
physics at the LHC looks promising.  Of course, more detailed
studies, also of the experimental design, will be necessary.

\section{Acknowledgment}
\label{Sec:Ackno}

It is a pleasure to thank K.  Eggert, H.  Gutbrod, A.  Morsch, S.
Sadovsky, and J.  Schukraft for their helpful, encouraging and
stimulating comments.

\appendix
\section*{The double equivalent photon approximation for inelastic
processes}

Our derivation for the equivalent photon approximation generalized to
inelastic processes follows in principle that of \cite{budnev75}
(using a slightly different definition of $\rho_{\mu\nu}$).  In this
short derivation we will not discuss the general applicability of the
EPA, but safely neglect the contribution of the longitudinal photons
in comparison to the transversal photons.  For a detailed discussion
of the applicability see, e.g., \cite{budnev75}.

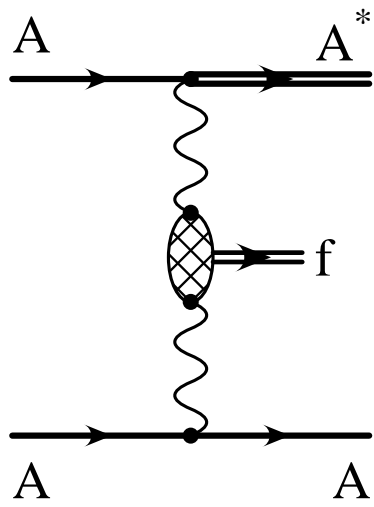
\begin{figure}
\caption{Inelastic-elastic process contributing to the
$\gamma$-$\gamma$ luminosity.  In the case of heavy ion collisions,
$A^*$ denotes an excited state of the nuclei, e.g., a GDR.}
\label{Fig:IEPA_GDR}
\end{figure}

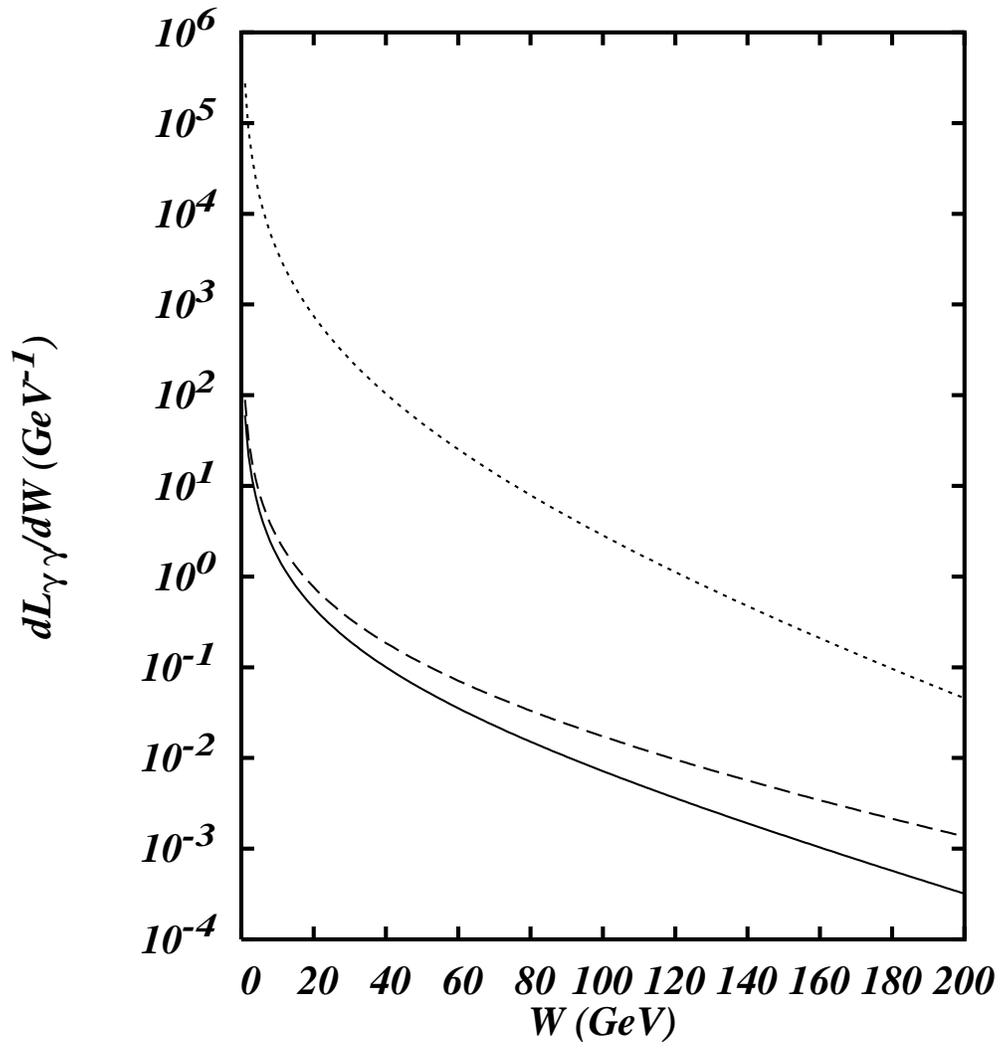
\begin{figure}
\caption{Comparison of the contributions of the elastic-elastic and
inelastic-elastic processes to the $\gamma$-$\gamma$ luminosity for a
Pb-Pb collision.  Shown are the results for elastic-elastic process
(dotted line) together with the inelastic-elastic processes using a
Gauss form factor (solid line) and a dipole form factor (dashed
line).}
\label{Fig:Lum_IEPA_Pb}
\end{figure}

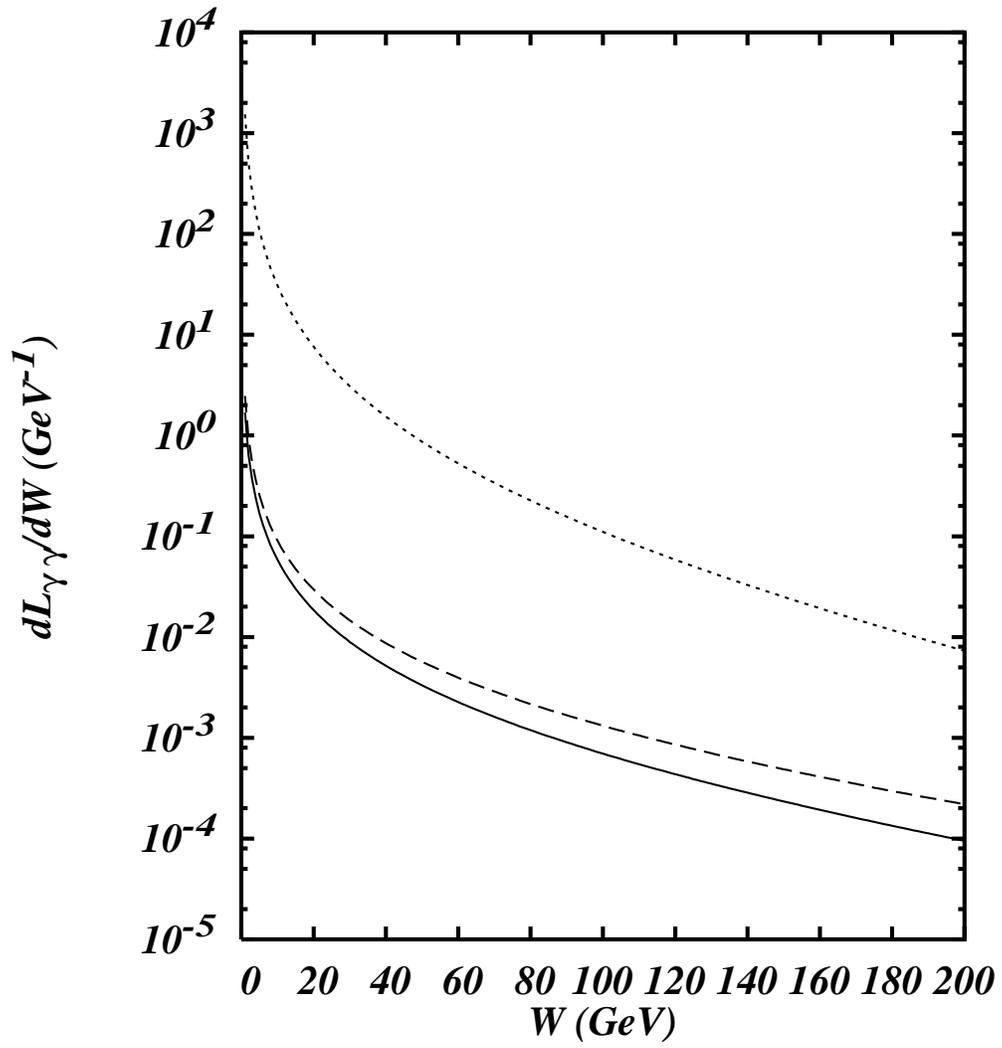
\begin{figure}
\caption{Same as Fig.~\protect\ref{Fig:Lum_IEPA_Pb} but now for a
Ca-Ca collision.}
\label{Fig:Lum_IEPA_Ca}
\end{figure}

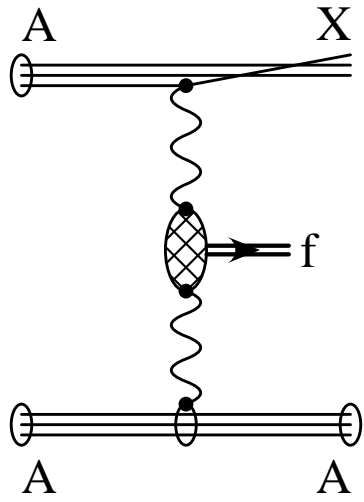
\begin{figure}
\caption{Inelastic Photon-photon process due to the photon emission
of an individual ``parton'' as opposed to the whole particle.  For
heavy ions we take the partons to be the protons, for $p$-$p$
collisions the quarks.}
\label{Fig:Dis}
\end{figure}

The cross section for the process (see also Fig.~\ref{Fig:IEPA_GDR})
is given by
\begin{equation}
d\sigma_{AA} = (2\pi)^4 \delta(P_f + P_1' - P_1 +P_2' - P_2)
\left| M \right|^2 \frac{1}{4I} \frac{d^3p_1'}{(2 \pi)^3 2 E_1'}
\frac{d^3p_2'}{(2 \pi)^3 2 E_2'} d\Gamma,
\end{equation}
where we do not make any assumptions on projectile and target and
also on the produced system $f$.  $P_f$ is the total momentum of this
system $f$, $d\Gamma$ its phase space. $|M|^2$ is given by
\begin{equation}
\left| M \right|^2 = \frac{1}{(q_1^2)^2} \frac{1}{(q_2^2)^2}
\Gamma_1^\mu \Gamma_1^{\mu' *} \Gamma_2^{\nu} \Gamma_2^{\nu' *}
W_{\mu\nu\mu'\nu'},
\end{equation}
with the momentum of the virtual photon $q_i = P_i' - P_i$ and the
electromagnetic transition current $\Gamma_i$.  In the EPA we neglect
the dependence of $W_{\mu\nu\mu'\nu'}$ on $q_i^2$ and replace it with
the one for real photons.  Therefore $W_{\mu\nu\mu'\nu'}$ depends
only on $\omega_1$ and $\omega_2$.  Averaging over the initial state
and summing over the final state, we get the ``photon densities''
$\rho_A$
\begin{equation}
\rho_A^{\mu\nu} := \frac{1}{2 J_i + 1} \sum_{M_i,M_f}
\Gamma^\mu \Gamma^{\nu*}.
\end{equation}
It is well known from general invariance considerations that the
general form of $\rho_A$ for an arbitrary transition has to be of the
form
\begin{equation}
\rho_A^{\mu\nu} =
\left(g^{\mu\nu} - \frac{q^\mu q^\nu}{q^2} \right) C  +
\left( P^\mu - \frac{qP}{q^2} q^\mu \right)
\left( P^\nu - \frac{qP}{q^2} q^\nu \right) D,
\end{equation}
where $C$ and $D$ are function of invariants only.

If the particle is moving relativistically, the photon momentum is
almost aligned to the beam axis.  In the EPA, we need only the
transverse components of $\rho$, transverse with respect to $q$.  In
the relativistic case, these transverse components are given by
\begin{equation}
\left< \rho_t^{\mu\nu} \right> = \left( - 2 C + D
\frac{q_\perp^2}{\omega^2} P^2 \right) \frac{1}{2}\sum_{\lambda}
\epsilon_\lambda^\mu \epsilon_\lambda^{\nu*},
\end{equation}
where $P=m\beta\gamma$ is the absolut value of the spatial components
of the heavy ion momentum and we have already averaged over the
transverse plane, which is normally not measured.  The
$\epsilon_\lambda$ are the two transverse polarizations of a
corresponding real photon.  We see that the photon density is
proportional to the unpolarized real photon density.  Using the
formula for the real photon-photon cross section, we can write the
cross section integrated over the scattered particles as
\begin{equation}
d\sigma_{AA} = \int \frac{d\omega_1}{\omega_1}
  \frac{d\omega_2}{\omega_2}
  N_1(\omega_1) N_2(\omega_2)
  d\sigma_{\gamma\gamma}(\omega_1,\omega_2),
\end{equation}
with the equivalent photon numbers $N_i(\omega_i)$, ($i=1,2$) given
by
\begin{equation}
N_i(\omega_i) = \int \frac{(-2 C_i
  + q_{i\perp}^2/\omega_i^2 P_{i}^2 D_i) \omega_i^2}
  {(2\pi)^3 2 E_i P_{i} (q_i^2)^2} d^2q_{i\perp}.
\end{equation}

For the elastic collision of a spinless particle we have $C=0$ and $D
= 4 \left| F \right|^2$, where $F$ is the elastic form factor.
$N(\omega)$ is then given by
\begin{equation}
N(\omega) = \int \frac{2 F^2(q^2) q_\perp^2}{(2\pi)^3
(\omega^2/\gamma^2 + q_\perp^2)^2} 2\pi q_\perp dq_\perp,
\label{Eq:iepa_el}
\end{equation}
where we have used the fact, that for an elastic collision $q^2 = -
(\omega^2/\gamma^2 + q_\perp^2)$.  Using a simple form factor $F^2
=4\pi Z^2 \alpha$ and integrating over $q_\perp$ from $0$ to a
cut-off value $\lambda=1/R$, we get the EPA spectrum in the leading
order as
\begin{equation}
N(\omega) = \frac{2 Z^2 \alpha}{\pi} \ln\left( \frac{\gamma}
{\omega R}\right).
\end{equation}
This is just the standard result \cite{budnev75,bertulani88}.

In order to find the corresponding expressions for $C$ and $D$ for a
photon emission with nuclear excitation, we work in the rest frame of
the initial nucleus and write $C$ and $D$ in invariant form.  In
order to distinguish between both systems, we write the photon
momentum in the rest frame as $(-\Delta,\vec k)$.  As the expressions
normally depend on $|\vec k|$, we define this as $k$, in contrast to
the Lorentz invariant $q$, which is given by $q^2 = \Delta^2 - k^2$.

Following deForest-Walecka \cite{deforest66}, we can write
\begin{eqnarray}
 \rho^{00} &=&
  4 \pi \left| M^C\right|^2\\
\rho^{\lambda\lambda'}
&=& \delta_{\lambda,\lambda'} 2 \pi \left(
 \left| T^e\right|^2 + \left| T^m\right|^2 \right)\\
\rho^{0 \lambda} &=& 0,
\end{eqnarray}
where $\lambda$, $\lambda'$ denotes directions transverse to $\vec
k$.  Together with the connection between $\Gamma^0$ and $\vec \Gamma
\vec k$ from the gauge invariance
\begin{equation}
\Delta \Gamma^0 = - \vec k \vec \Gamma,
\end{equation}
this determines all components of the electromagnetic tensor.

Using these equations, $C$ and $D$ can be expressed in terms of
$T^e$, $T^m$, and $M^C$ \cite{walecka83}.  We get
\begin{eqnarray}
C &=& - 2 \pi \left[ |T^e|^2 + |T^m|^2 \right] \\
D &=& \frac{(q^2)^2}{k^4 M^2} 2 \pi \left[2 |M^C|^2 -
\frac{k^2}{q^2}\left( |T^e|^2 + |T^m|^2 \right)  \right].
\label{Eq:iepa_CD}
\end{eqnarray}

As a case of special importance, we apply this to the GDR excitation.
In the Goldhaber-Teller model of the GDR as a harmonic oscillator
\cite{deforest66,goldhaber48}, we get
\begin{eqnarray}
\left| M^C \right|^2 &=&
 4 M^2 \left(\frac{N}{A}\right)^2 \frac{k^2}{2 \mu}
\frac{1}{\Delta} \frac{F^2(k)}{4\pi},\\
\left| T^e \right|^2 &=&
8 M^2 \left(\frac{N}{A}\right)^2 \frac{k^2}{2 \mu} \frac{1}{\Delta}
\left(\frac{\Delta}{k}\right)^2
\frac{F^2(k)}{4\pi},
\end{eqnarray}
where $M$ is the mass of the nucleus and $\mu = (N Z / A) M$.
Please note that the form factor $F$ appearing in these equations is
the elastic form factor of the nuclei as a function of $k^2 =
\Delta^2 - q^2$.  Only in the elastic case are $k^2$ and $q^2$
identical.  Using Eq.~(\ref{Eq:iepa_CD}) we get
\begin{eqnarray}
C &=& - 4 M^2 \left(\frac{N}{A}\right)^2 \frac{F^2}{2 \mu} \Delta,\\
D &=&   4 \left(\frac{N}{A}\right)^2 \frac{F^2}{2 \mu}
\frac{-q^2}{\Delta}.
\end{eqnarray}
The EPA spectrum is then given by
\begin{equation}
N(\omega) = \int \left(\frac{N}{A} \right)^2 \frac{4 F^2}{2\mu}
\frac{2 M^2 \Delta^2 \omega^2 + q_\perp^2 P^2 |q|^2}{(2\pi)^3 2 E P
(q^2)^2 \Delta} d^2q_\perp.
\label{Eq:iepa_inel}
\end{equation}

In the elastic case, $q^2$ was given by $-(\omega^2/\gamma^2 +
q_\perp^2)$.  For the inelastic case, we get for $\gamma \gg 1$
\begin{eqnarray}
q^2 &=& \Delta^2
  - \left( \frac{\omega}{\gamma \beta}
  + \frac{\Delta}{\beta}\right)^2
  - q_\perp^2 \\
&\approx& - \left[\frac{\omega}{\gamma}
  \left(\frac{\omega}{\gamma} + 2 \Delta\right) + q_\perp^2 \right].
\end{eqnarray}

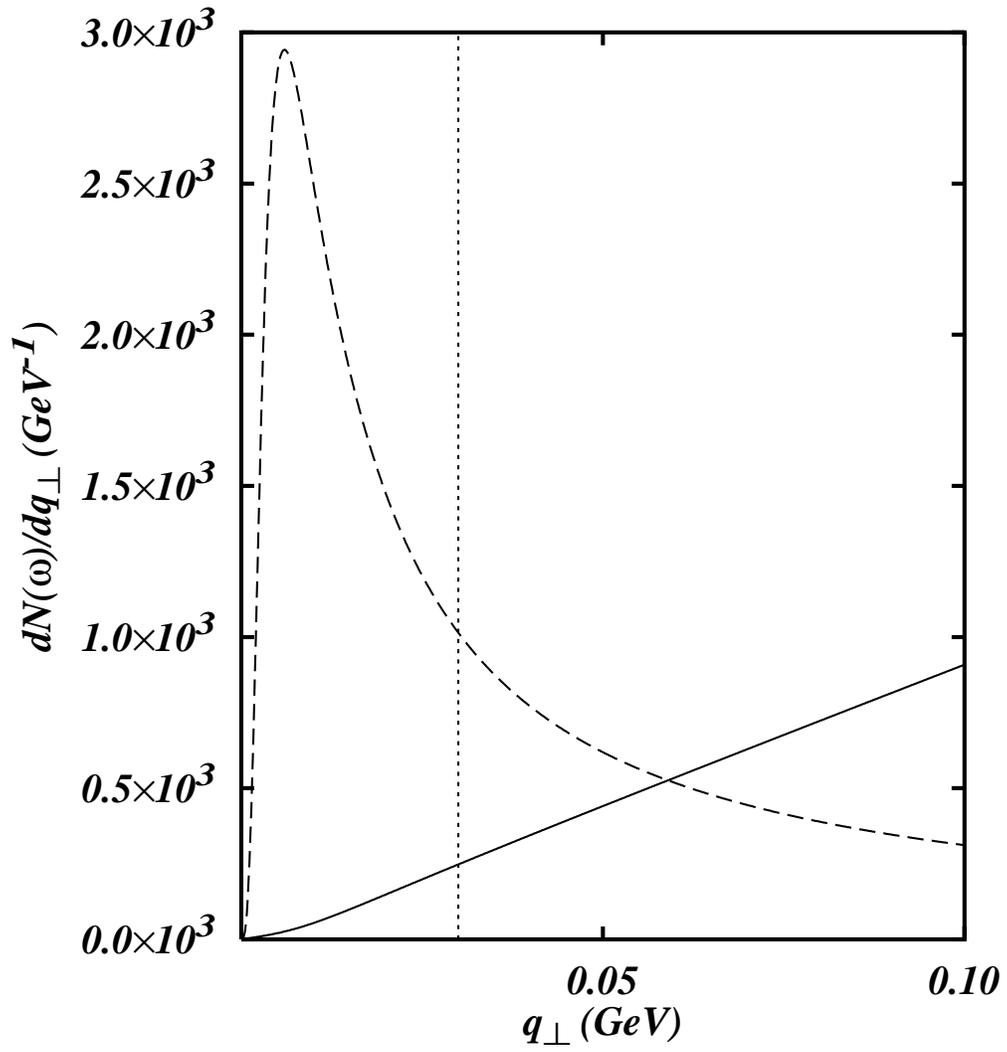
\begin{figure}
\caption{The differential photon number $dN(\omega)/dq_\perp$ is
compared for a Pb-Pb collision and $\omega=10$~GeV for elastic
(dotted line, Eq.~(\protect\ref{Eq:iepa_el})) and inelastic photon
emission (solid line, Eq.~(\protect\ref{Eq:iepa_inel})).  The
vertical line shows the range of $q_\perp$, which is limited due to
the form factor.  The result of the inelastic case has been
multiplied by a factor of 400.}
 \label{Fig:dndqp}
\end{figure}

The form factor $F$ appearing in the EPA spectrum is the elastic form
factor of the nucleus, written as a function of $k^2 = \Delta^2
-q^2$.  In the elastic case, the spectrum was insensitive to the
detailed form of the form factor.  Comparing the integrand of
Eq.~(\ref{Eq:iepa_el}) and Eq.~(\ref{Eq:iepa_inel}) without a form
factor, we see, that for inelastic processes $dN/dq_\perp$ increases
linearly for large $q_\perp$, contrary to the elastic case.  In
Fig.~\ref{Fig:dndqp} we compare both in the case of a Pb nucleus.
Therefore the photon spectrum is sensitive to the form factor in the
inelastic case.  In our calculations we use a dipole form factor
\begin{equation}
F = \Lambda^4 / (\Lambda^2 + k^2)^2,
\end{equation}
and a Gaussian form factor
\begin{equation}
F = \exp\left(-\frac{k^2}{\lambda^2}\right),
\end{equation}
where the parameters have been chosen to get the right root mean
square radius of the nuclei.  Integrating over the transverse
momentum we get $N(\omega)$ which has been used in order to get the
elastic-inelastic $\gamma$-$\gamma$ luminosity $dL/dW$.

\narrowtext
\begin{table}[tbp]
\begin{center}
\begin{tabular}{lcrr}
Collider & Luminosity ($cm^{-2} s^{-1}$) & $E_{cm}$ & $\gamma$\\
\hline
LEP200      & $1 \times 10^{32}$ & 100 GeV & $2 \times 10^5$\\
$p$-$p$ at LHC & $1 \times 10^{33}$ & 7 TeV & 7500\\
Pb-Pb at LHC & $5 \times 10^{26}$ & 7 $A$TeV $Z/A$ & 2900 \\
Ca-Ca at LHC & $5 \times 10^{30}$ & 7 $A$TeV $Z/A$ & 3700 \\
NLC/PLC ($z$=0cm) & $2 \times 10^{33}$ & 500 GeV & \\
\end{tabular}
\end{center}
\caption{The collider parameters used in the comparison of the
different $\gamma$-$\gamma$ luminosities.  Please see text for
References of the different colliders and explanation of $z$.}
\label{Tab:Collider}
\end{table}
\end{document}